\documentclass[10pt, prd, twocolumn, showpacs]{revtex4}

\usepackage{amsmath}
\usepackage{amssymb}
\usepackage{graphicx}
\usepackage{bm} 

\newcommand{\al}[0]{\alpha} 
\newcommand{\be}[0]{\beta}
\newcommand{\de}[0]{\delta}
 
\newcommand{\dd}[0]{\partial}

\begin{document}

\title{Entropy of a self-gravitating 
electrically charged thin shell
and the  black hole limit}
\author{Jos\'{e} P. S. Lemos}
\email{joselemos@ist.utl.pt}
\affiliation{Centro Multidisciplinar de Astrof\'{\i}sica, CENTRA,
Departamento de F\'{\i}sica, Instituto Superior T\'ecnico - IST,
Universidade de Lisboa - UL, Avenida Rovisco Pais 1,
1049-001 Lisboa, Portugal\,\,\,}
\author{Gon\c{c}alo M. Quinta}
\email{goncalo.quinta@ist.utl.pt}
\affiliation{Centro Multidisciplinar de Astrof\'{\i}sica, CENTRA,
Departamento de F\'{\i}sica, Instituto Superior T\'ecnico - IST,
Universidade de Lisboa - UL, Avenida Rovisco Pais 1,
1049-001 Lisboa, Portugal\,\,}
\author{Oleg B. Zaslavskii}
\email{ozaslav@kharkov.ua}
\affiliation{ Department of Physics and Technology, Kharkov V.
N. Karazin National University, 4 Svoboda Square, Kharkov 61022,
Ukraine, and Institute of Mathematics and Mechanics, Kazan Federal
University, 18 Kremlyovskaya St., Kazan 420008, Russia}

\begin{abstract}

A static self-gravitating electrically charged spherical thin shell
embedded in a (3+1)-dimensional spacetime is used to study the
thermodynamic and entropic properties of the corresponding spacetime.
Inside the shell, the spacetime is flat, whereas outside it is a
Reissner-Nordstr\"om spacetime, and this is enough to establish the
energy density, the pressure, and the electric charge in the shell.
Imposing that the shell is at a given local temperature and that the
first law of thermodynamics holds on the shell one can find the
integrability conditions for the temperature and for the thermodynamic
electric potential, the thermodynamic equilibrium states, and the
thermodynamic stability conditions.  Through the integrability
conditions and the first law of thermodynamics an expression for
the shell's entropy can be calculated.  It is found that the shell's
entropy is generically a function of the shell's gravitational and 
Cauchy radii alone.  A plethora of sets of temperature and electric
potential equations of state can be given.  One set of equations of
state is related to the Hawking temperature and a precisely given
electric potential. Then, as one pushes the shell to its own
gravitational radius and the temperature is set precisely equal to the
Hawking temperature, so that there is a finite quantum backreaction
that does not destroy the shell, one finds that the entropy of the
shell equals the Bekenstein-Hawking entropy for a black hole.  The
other set of equations of state is such that the temperature is
essentially a power law in the inverse Arnowitt-Deser-Misner
(ADM) mass and the electric
potential is a power law in the electric charge
and in the inverse
ADM mass. In this case, the
equations of thermodynamic stability are analyzed, resulting in
certain allowed regions for the parameters entering the problem. Other
sets of equations of state can be proposed.  Whatever the initial
equation of state for the temperature, as the shell radius approaches
its own gravitational radius, the quantum backreaction imposes the
Hawking temperature for the shell in this limit.  Thus, when the
shell's radius is sent to the shell's own gravitational radius the
formalism developed allows one to find the precise form of the
Bekenstein-Hawking entropy of
the correlated black hole.

\end{abstract}

\maketitle


\section{Introduction}

In general relativity, in 3+1 dimensions, a 
black hole spacetime is characterized by its
conserved charges and the fundamental constants.  The conserved
charges are for example the Arnowitt-Deser-Misner
(ADM) mass $m$ and the electric charge $Q$.
The fundamental constants are the two constants of the theory, namely,
the gravitational constant
$G$, and the velocity of light
(which is set to one). 
In an analysis of quantum aspects of a black hole, such as
the black hole entropy
and its inherent
degrees of freedom, the other fundamental constant
in physics, Planck's constant 
$\hbar$, also appears naturally.
With these three constants one makes the
Planck length $l_p=\sqrt{G\hbar}$ 
and the Planck area $A_p=l_p^2$. Also, $m$, $Q$,
$G$, and the velocity of light give the horizon
radius $r_+$ and so the horizon area $A_+=4\pi r_+^2$.  Then, the
Bekenstein-Hawking entropy of a black hole, 
given by $S_{\rm
bh}=\frac14\,\frac{A_+}{A_p}$
\cite{beken,hawk1,hawk2}, where the Boltzmann constant is set to
one, is a measure of how many Planck areas there are in the horizon
area. It also shows that black hole quantum mechanics, and
consequently black hole entropy, is in its essence and generality a
process of pure quantum gravity, as no other constants besides the
gravitational constant $G$, the velocity of light, and Planck's constant
$\hbar$ enter, through 
the Planck area ${A_p}$, 
in the final process. In addition, it suggests that the
ultimate degrees of freedom that inhabit the realm of quantum gravity
are in the area of the enclosing region,
rather than within the volume
as it is the case for ordinary matter
\cite{thooft,suss} (for a review see, e.g., \cite{lemos}).  
However, since there is no
quantum gravity theory at hand,
black hole entropy is still an enigma although
there has been progress in its understanding, 
especially through the resort to gravitational 
low-energy quantum theories.

Since black holes are vacuum solutions, and our primitive concepts of
entropy are based on the quantum properties of matter, it would be
useful to have a spacetime with matter and study its thermodynamic and
entropic properties. One then can look for a limit where a black hole
might emerge. In this way, one can have hints to how a black hole's
entropy develops.  We are thus interested in a system which contains
both gravitational and material degrees of freedom but which does not
introduce too many complexities due to the matter constitution.

The next simplest solution to a black hole solution, is a vacuum
solution except for an infinitesimally thin region of spacetime where
there is matter, i.e., a self-gravitating thin shell.  As a thin shell
is the nearest to a vacuum solution one can have, it is a very useful
system that allows one to probe almost pure spacetime properties.  A
thin shell is defined as an infinitesimally thin surface which
partitions spacetime into an interior region and an exterior
region. Since it corresponds to some sort of matter and the spacetime
properties must reflect it, the thin shell should satisfy some
conditions in order for the entire spacetime to be a valid solution of
the Einstein equations. Such conditions relate the stress-energy
tensor of the shell to the extrinsic curvature of the spacetime. The
stress-energy tensor yields the density and pressure, and in general
the matter properties are also the equations of state, such as the
temperature and possibly others, and the entropy.  A particularly
simple thin shell is one that is static and is spherically symmetric.

Suppose then a self-gravitating static spherical thin shell.  Assume the
simplest case, the inner spacetime is Minkowski and the outer
spacetime is Schwarzschild.  One can then work out its dynamics and
thermodynamic properties, such as the temperature and entropy.  In an
elegant work, by finding the surface energy density and pressure, and
imposing that the shell is at a given local temperature $T$, and so
using a canonical ensemble, Martinez \cite{Mart} found those
thermodynamic 
properties for the simplest shell, characterized by its rest mass $M$
and radius $R$. In \cite{Mart} only shells whose matter obeyed the
dominant energy condition, and so the radius $R$ greater than a given
value, were considered.  Martinez'z approach \cite{Mart} draws in many
respects from York's work \cite{york1} where the thermodynamic
properties of a pure Schwarzschild black hole is treated using a
canonical ensemble, i.e., imposing a fixed temperature on some
fictitious massless
shell at a definite radius outside the event horizon.
Another reason that motivates the use of thin shells is the fact that
they can be taken with some ease to their own gravitational radius,
i.e., to the black hole limit.  If one does that, as was done in
\cite{lemoszaslavskii}, one recovers the black hole entropy, 
i.e., the entropy $S$ of the shell at its
own gravitational radius is $S=S_{\rm
bh}=\frac14\,\frac{A_+}{A_p}$ for such a matter configuration at the
black hole limit. Such a configuration is called a quasiblack hole.
Thus, the black hole thermodynamic properties can be studied by a
direct computation if thin shells are used.

It is important to generalize Martinez's work \cite{Mart} for
electrically charged shells, which we will do here.  We consider that
the shell has an electric charge $Q$.  In this case, the inner
spacetime is Minkowski and the outer spacetime is
Reissner-Nordstr\"om.  One can then work out the shell's dynamics and
thermodynamic properties, such as the energy density, the pressure,
the electric potential temperature, and the entropy.  Due to the
introduction of a new state variable in the thermodynamic system,
namely, the additional thermodynamic electric potential, the
calculations become considerably more complex. At the same time the
richness of the physical results increases as well.  We take the shell
to its own gravitational radius, the black hole end point, which is
meaningful in the calculation of the shell's entropy $S$, and find
that the entropy is equal to the Bekenstein-Hawking entropy.  The
extremal $\sqrt{G}m= Q$ limit can then be taken which gives the same
expression for the entropy, i.e., $S=S_{\rm
bh}=\frac14\,\frac{A_+}{A_p}$, see, however,
\cite{pretvolisr,lemoszaslaextremal} for a discussion of the entropy
of extremal black holes taken from extremal shells.  Electrically
charged black holes were studied in \cite{yorketal,pecalemos}, where
the thermodynamic properties of a pure Reissner-Nordstr\"om black hole
are treated using a grand canonical ensemble, i.e., imposing a fixed
temperature and electric potential at some definite radius outside the
event horizon.

There are other works that used thin shells to understand the
thermodynamics and the evolution of the entropy in certain spacetimes.
In \cite{lemosquinta1,lemosquinta2} the formalism of Martinez
\cite{Mart} was used to study three-dimensional thin shells including
the thermodynamics of a thin shell with a 
static Ba\~nados-Teitelboim-Zanelli (BTZ) outer spacetime.  In
\cite{daviesfordpage,hiscock} thin shells with a black hole inside
were used to understand how the entropy of the spacetime 
evolves as the shell
approaches its own event horizon.

We analyze static thin shells
using the junction condition formalism 
established in \cite{Israel} with the complement to electrically
charged shells 
developed 
in \cite{Kuchar}. Our thermodynamic approach, follows
the general approach for thermodynamic systems given in \cite{callen},
as does the approach of \cite{Mart}.

We will adopt the following line of work.  In Sec.~\ref{thinsh} we
study a static spherical symmetric thin shell whose interior is
Minkowski and exterior is Reissner-Nordstr\"om. We find the main
properties of the global spacetime as well as the rest energy density,
and thus the rest mass, the pressure in the shell, and the shell's
electric charge.  In Sec.~\ref{thermo} we exhibit the first law of
thermodynamics, find the generic integrability conditions and the
stability conditions.  In Sec.~\ref{eqsos} we use the spherical shell
whose dynamics 
is displayed in Sec.~\ref{thinsh}. 
We present the three independent 
thermodynamic variables $(M,R,Q)$
and then through the integrability conditions find the functional
dependence for the temperature $T$ and the thermodynamic electric
potential $\Phi$
on those variables.  Then in Sec.~\ref{entro} the 
differential for the entropy $S$ of the shell
is obtained as a differential on the gravitational
radius $r_+$ and the Cauchy horizon radius $r_-$
and up to two functions which depend on 
$r_+$ and $r_-$. Those
functions are essentially the inverse of the temperature 
and the electric potential of the shell if it were
located at infinity. Moreover, it is shown 
that the two functions are related by a specific differential
equation, and 
that the entropy of the shell is a function of $r_+$ and $r_-$
alone, which themselves are functions
of $(M,R,Q)$.  In Sec.~\ref{bhlimit}, to advance further, one needs to
specify the form of the equations of state. We give a
particular set of 
equations of state that will lead us with some ease to the
black hole entropy when the shell is taken to its own gravitational
radius. Indeed, 
by choosing the Hawking temperature due to quantum-mechanical
arguments and a precise electric potential, 
the entropy of a charged black hole will naturally emerge.
We then compare our approach with the usual thermodynamic approach for
black holes.  In Sec.~\ref{eqstate} we give another
simple set of phenomenological 
equations of
state for the temperature and the electric
potential, where
free parameters encoding the details of the matter fields will
naturally appear. This set of equations of state
allows us to 
also find the entropy and study 
analytically the stability
conditions.  In Sec.~\ref{other}
we briefly discuss other interesting equations of state.
Finally, in Sec.~\ref{conc} we conclude.  We
leave for Appendix A a study of the dominant energy condition 
of the matter fields in the shell which is
not important in the thermodynamic study, but which is interesting to
have.  In Appendix B we derive the equations of thermodynamic
stability for a system with three independent variables.

\section{The thin-shell spacetime}
\label{thinsh}

\subsection{The Einstein-Maxwell equations}
\label{emeqs}

We start with the Einstein-Maxwell equations in 3+1 dimensions
\begin{equation}
G_{\al\be}=8\pi G\,
T_{\al\be}\,, 
\label{ein} 
\end{equation} 
\begin{equation}
\nabla_{\beta}F^{\alpha\beta}=4\pi J^{\alpha}\,.
\label{max} 
\end{equation} 
$G_{\al\be}$ is the Einstein tensor, built from the spacetime metric
$g_{\al\be}$ and its first and second derivatives, 
$8\pi G$ is the coupling, with $G$ being the gravitational constant in
3+1 dimensions and we 
are using units in which the velocity
of light is one, and $T_{\al\be}$ is the energy-momentum tensor. 
$F_{\alpha\beta}$ is the
Faraday-Maxwell tensor,
$J_\alpha$ is the electromagnetic four-current and $\nabla_\beta$
denotes covariant derivative.  
The other Maxwell equation $\nabla_{[\gamma}F_{\alpha\beta]}=0$,
where $[...]$ means antisymmetrization, 
is automatically satisfied for a properly defined
$F_{\alpha\beta}$.
Greek indices
will be used for spacetime indices and run as $\alpha,\beta=0,1,2,3$,
with 0 being the time index.

\subsection{The thin-shell gravitational junction conditions}
\label{gravj}

We consider now a two-dimensional timelike 
massive electrically charged shell with radius $R$,
which we will call $\Sigma$.  The shell partitions spacetime into two
parts, an inner region $\mathcal{V}_i$ and an outer region
$\mathcal{V}_o$. 
In order to find a global spacetime solution for the
Einstein equation, Eq.~(\ref{ein}), we will use the thin-shell
formalism developed in \cite{Israel}.

First, we  specify the metrics on each side of the shell. 
In
the inner region $\mathcal{V}_i$ ($r< R$) we assume the spacetime
is flat, i.e.
\begin{align}\label{LEI}
ds_i^2  & = g_{\al\be}^i dx^\alpha dx^\beta =\nonumber\\&
-dt_i^2 + dr^2 + r^2\, d\Omega^2\,,\quad r< R\,,
\end{align}
where $t_i$ is the inner time coordinate,
polar coordinates 
$(r,\theta,\phi)$ are used, and 
$d\Omega^2 = d\theta^2 + \sin^2\theta\,d\phi^2$.
In the outer region $\mathcal{V}_o$ ($r> R$), 
the spacetime is described by the
Reissner-Nordstr\"{o}m line element
\begin{align}\label{LEO}
ds_o^2 & = g_{\al\be}^o dx^\alpha dx^\beta = \nonumber\\
& \hspace{-3mm} -\left(1 - \frac{2Gm}{r} + \frac{G Q^2}{r^2}\right) 
dt_o^2 + \frac{dr^2}{1 - \dfrac{2Gm}{r} + \dfrac{G Q^2}{r^2}} \nonumber \\
& \hspace{7mm} + r^2 d\Omega^2 \,, \quad r> R\,,
\end{align}
where $t_o$ is the outer time coordinate, and again
$(r,\theta,\phi)$ are polar coordinates, and
$d\Omega^2 = d\theta^2 + \sin^2\theta\,d\phi^2$.
The constant $m$ is to be  interpreted as  
the ADM mass, or energy, and
$Q$ as the electric charge. 
Finally, 
on the hypersurface itself, $r=R$,
the metric 
$h_{ab}$ is that of a 2-sphere with an additional time dimension,
such that,
\begin{equation}
ds_{\Sigma}^2 = h_{ab} dy^a dy^b =
-d\tau^2 + R^2(\tau) d\Omega^2\,, \quad r= R\,,
\label{intrinsmetr}
\end{equation}
where we have chosen $y^a=(\tau,\theta,\phi)$ as 
the time and spatial coordinates on the
shell. We have adopted the convention to use latin indices for the
components on the hypersurface. The time coordinate $\tau$ is the
proper time for an observer located at the shell. The shell radius is
given by the parametric equation $R= R(\tau)$ for an observer on the
shell. On each side of the hypersurface, the parametric equations for
the time and radial
coordinates are denoted by 
$t_{i}=T_{i}(\tau)$, $r_{i}=R_{i}(\tau)$,
and $t_{o}=T_{o}(\tau)$, $r_{o}=R_{o}(\tau)$.
The metric $h_{ab}$ is also called the
induced metric and can be written in terms of the 3+1-dimensional
spacetime metric $g_{\al\be}$. In particular, viewed from each side of
the shell, the induced metric is given by
\begin{equation}
h^{i}_{ab} = g^{i}_{\al\be} \, e^{\al}_{i}{}_a \,
e^{\be}_{i}{}_b\,,
\quad
h^{o}_{ab} = g^{o}_{\al\be} \, e^{\al}_{o}{}_a \,
e^{\be}_{o}{}_b\,,
\end{equation}
where $e^{\al}_{i}{}_a$ 
and $e^{\al}_{o}{}_a$ are tangent vectors to the hypersurface
viewed from the inner and outer regions, respectively. 
With these last expressions, we have all
the necessary information to employ the formalism developed in
\cite{Israel}. 
We will also apply this formalism to electrically charged
systems which was displayed first in
\cite{Kuchar}.

The thin-shell formalism states that two junction conditions are
needed in order to have a smooth change across the
hypersurface. The first junction condition is expressed by the
relation
\begin{equation}
[h_{ab}]=0\,,
\end{equation}
where the parentheses symbolize the jump in the quantity across the
hypersurface, which in this case is the induced metric. This condition
immediately implies that $h^{i}_{ab}=h^{o}_{ab}=h_{ab}$, or explicitly
\begin{align} \label{J1}
& -\left(1 - \frac{2Gm}{r} + \frac{G Q^2}{r^2}\right) \dot{T}_o^2 +
\nonumber \\ & \hspace{8mm} \frac{\dot{R}_o^2}{\left(1 -
\dfrac{2Gm}{r} + \dfrac{G Q^2}{r^2}\right)} = -\dot{T}_i^2 +
\dot{R}_i^2 = -1\,,
\end{align}
where a dot denotes differentiation with respect to $\tau$. 
The second junction condition is related to the 
inner and outer extrinsic curvature
$K^a_{i}{}_{b}$ 
and $K^a_{o}{}_{b}$, respectively,
defined as
\begin{equation}
K^a_{i}{}_{b} = \left(\nabla_{\be}\, n^{i}_{\al}\right) \, e^{\al}_{i}{}_c 
\, e^{\be}_{i}{}_b \, h^{ca}_{i}\,,\,
K^a_{o}{}_{b} = \left(\nabla_{\be}\, n^{o}_{\al}\right) \, e^{\al}_{o}{}_c 
\, e^{\be}_{o}{}_b \, h^{ca}_{o}
\,,
\label{extr1}
\end{equation}
where 
$n^{i}_\alpha$ and $n^{o}_\alpha$,
are the inner and outer normals to the shell,
respectively. The second junction condition
then says $\left[K^{a}{}_{b}\right]=0$ if the metric is to be smooth
across the hypersurface. However, this condition can be violated, in
which case it can be physically interpreted as the existence of a thin
matter shell where the hypersurface is located. In addition, the
shell's stress-energy tensor $S^{a}{}_{b}$ is related to the jump in
the extrinsic curvature through the Lanczos equation, namely,
\begin{equation}
S^{a}{}_{b}=-\frac{1}{8 \pi G}
\left([K^{a}{}_{b}]-[K]h^{a}{}_{b}\right)\,,
\label{extr2}
\end{equation}
where $K = h^{b}{}_{a} K^{a}{}_{b}$. Proceeding then to the
calculation of the extrinsic curvature components, 
one can show that they are given by the general expressions
\begin{align}
K^{\tau}_i{}_{\tau} &= \frac{\ddot{R}}{\sqrt{1+\dot{R}^2}}\,, 
\label{a1}\\
K^{\tau}_o{}_{\tau} &= \frac{-\frac{G \dot{m}}{R\,\dot{R}}-\frac{G
Q^2}{R^3}+\frac{G m}{R^2}+\ddot{R}}{\sqrt{1-\frac{2Gm}{R} + \frac{G
Q^2}{R^2}+\dot{R}^2}}\,, \label{a2}\\
K^{\phi}_i{}_{\phi} = K^{\theta}_i{}_{\theta} &=
\frac{1}{R}\sqrt{1+\dot{R}^2}\,,\label{a3}\\
K^{\phi}_o{}_{\phi} = K^{\theta}_o{}_{\theta} &=
\frac{1}{R}\sqrt{1-\frac{2Gm}{R} + \frac{G Q^2}{R^2}+\dot{R}^2}\,.
\label{a4}
\end{align}
Using Eqs.~(\ref{a1})-(\ref{a4}) in Eq.~(\ref{extr2}), one can
calculate the non-null components of the stress-energy tensor $S_{ab}$
of the shell. In particular, we will assume a static shell, such that
$\dot{R}=0$, $\ddot{R}=0$, and $\dot{m}=0$. In that case, we are led to
\begin{align}
S^{\tau}{}_{\tau} & = \frac{\sqrt{1-\frac{2Gm}{R} + \frac{G Q^2}{R^2}}
- 1}{4 \pi G R}\,, \label{S1} \\
S^{\phi}{}_{\phi} = S^{\theta}{}_{\theta} &  = 
\frac{\sqrt{1-\frac{2Gm}{R}+\frac{G Q^2}{R^2}}-1}{8 \pi G R} + 
\nonumber\\
&
\frac{\frac{m G}{R}-\frac{G Q^2}{R^2}}{8 \pi G R
\sqrt{1-\frac{2Gm}{R}+\frac{G Q^2}{R^2}}} \label{S2}\,.
\end{align} 
To
further advance, one needs to specify what kind of matter the shell
is made of, which we will consider to be a perfect fluid with surface
energy density $\sigma$ and pressure $p$. This implies that the
stress-energy tensor will be of the form
\begin{equation}
S^{a}{}_{b} = (\sigma +p) u^a u_b + p
h^{a}{}_{b}\,,
\label{perffluid}
\end{equation}
where 
$u^a$ is the three-velocity of a shell element. We thus find that
\begin{equation}
S^{\tau}{}_{\tau} = -\sigma\,,
\label{lam}
\end{equation}
\begin{equation}
S^{\theta}{}_{\theta} = S^{\phi}{}_{\phi} = p
\label{press}\,.
\end{equation}
Combining Eqs.~(\ref{lam})-(\ref{press})
with Eqs.~(\ref{S1})-(\ref{S2}) results in the equations
\begin{align}
\sigma = & \frac{1-\sqrt{1-\frac{2Gm}{R} + \frac{G Q^2}{R^2}}}{4 \pi G
R}\,, \label{SS1} \\ 
p = & \frac{\sqrt{1-\frac{2Gm}{R}+\frac{G
Q^2}{R^2}}-1}{8 \pi G R} + \nonumber \\
&
\frac{\frac{m G}{R}-\frac{G Q^2}{R^2}}{8
\pi G R \sqrt{1-\frac{2Gm}{R}+\frac{G Q^2}{R^2}}}
\label{SS2}\,.
\end{align} 
Note that Eq.~(\ref{SS2}) is purely a
consequence of the Einstein equation which is encoded in the junction
conditions. Thus, although no information about the matter fields of
the shell has been given, we know that they must have
a pressure equation of the form
(\ref{SS2}), otherwise no mechanical equilibrium can be achieved.

It is useful to define the shell's redshift 
function $k$ as
\begin{equation}
k=\sqrt{
1 - \frac{2Gm}{R} + \frac{G Q^2}{R^2}}
\label{red0}\,.
\end{equation}
Equation~(\ref{red0})
allows Eqs.~(\ref{SS1})-(\ref{SS2}) to be written as
\begin{align} 
\sigma = & \frac{1-k}{4\pi G R} \label{Sig1}\,, \\
p = & \frac{R^2(1-k)^2 - GQ^2}{16 \pi G R^3 k}\,. 
\label{pQk1}
\end{align}
From the energy density $\sigma$ of the shell we can define 
the rest mass $M$ through the equation
\begin{equation}
\sigma = \frac{M}{4\pi R^2} \,.
\label{sigmarestmasssigma}
\end{equation}
Note that from Eqs.~(\ref{Sig1}) and (\ref{sigmarestmasssigma})
one has 
\begin{equation}\label{M1}
M = \frac{R}{G}(1-k).
\end{equation}
Using Eqs.~(\ref{red0}) and (\ref{M1}), we are led to
an equation for the ADM mass $m$,
\begin{equation}\label{m0}
m = M - \frac{G M^2}{2R} + \frac{Q^2}{2R}\,.
\end{equation}
This equation is intuitive on physical grounds as it states that the
total energy $m$
of the shell is given by its mass $M$ minus the energy
required to built it against the action of gravitational and
electrostatic forces,
i.e., $- \frac{G M^2}{2R} + \frac{Q^2}{2R}$. 
For $Q=0$, we recover the result derived in
\cite{Mart}.
Note that Eq.~(\ref{m0}) is also purely a
consequence of the Einstein equation encoded in the junction
conditions, i.e., although no information about the matter fields of
the shell has been given, we know that they must have an ADM mass
given by Eq.~(\ref{m0}).

The gravitational radius $r_+$ and
the Cauchy horizon $r_-$ of the shell
spacetime are given by the zeros 
of the $g_{00}^o$ in Eq.~(\ref{LEO}). They are  
then
\begin{equation}
r_{+} = G\,m +\sqrt{G^2m^2 - G Q^2}\,,
\label{horradi}
\end{equation}
\begin{equation}
r_{-} = G\,m - \sqrt{G^2m^2 - G Q^2}\,,
\label{horradicauch}
\end{equation}
respectively.
The gravitational radius $r_+$ is also the horizon radius when the
shell radius $R$ is inside $r_+$, i.e., the spacetime contains a black
hole. Although they have the same expression, conceptually, the
gravitational and horizon radii are distinct.  Indeed, the
gravitational radius is a property of the spacetime and matter,
independently of whether there is a black hole or not. On the other
hand, the horizon radius exists only when there is a black hole.

The gravitational radius $r_+$ 
and the Cauchy horizon $r_-$
in Eqs.~(\ref{horradi})-(\ref{horradicauch})
can be inverted to give
\begin{equation}
m=\frac{1}{2G}\left( r_{+} +  r_{-}\right)\,,
\label{invhorradi}
\end{equation}
\begin{equation}
Q=\sqrt{
\frac{
r_{+}r_{-}}{G}
}\,.
\label{invhorradicauch}
\end{equation}
From Eq.~(\ref{horradi}) one can define 
the gravitational area $A_+$ as
\begin{equation}\label{arear+}
A_+=4\pi\,r_+^2\,.
\end{equation}
This is also the event horizon area when there is
a black hole.
Using Eqs.~(\ref{horradi})-(\ref{horradicauch})
implies that $k$ in Eq.~(\ref{red0}) can be written as
\begin{equation}\label{red}
k=\sqrt{\Big(1-\frac{r_+}{R}\Big)\Big(1-\frac{r_-}{R}\Big)}\,.
\end{equation}

The area $A$ of the shell, an important quantity, is from 
Eq.~(\ref{intrinsmetr}) given by 
\begin{equation}
A=4 \pi R^2\,.
\label{area1}
\end{equation}

\subsection{The thin-shell electromagnetic junction conditions}
\label{emj}

Now we have to deal with Eq.~(\ref{max}).
The
Faraday-Maxwell tensor $F_{\alpha\beta}$ is usually defined in terms
of an electromagnetic four-potential $A_\alpha$ by
\begin{equation}
F_{\al\be} = \partial_{\al}A_{\be}-\partial_{\be}A_{\al}\,,
\label{empot}
\end{equation}
where $\partial_{\beta}$ denotes partial
derivative.

To use the thin-shell formalism related to the electric part
we need to specify the vector potential $A_\alpha$
on each side of the shell.
We assume an electric ansatz for the 
electromagnetic four-potential $A_\alpha$, i.e., 
\begin{equation}
A_{\alpha} = (-\phi,
0,0,0)\,, 
\label{empotspec}
\end{equation}
where $\phi$ is thus the electric potential.
In
the inner region $\mathcal{V}_i$ ($r < R$) the spacetime
is flat. 
So the Maxwell equation 
$\nabla_{\beta}F^{\alpha\beta}=\frac{1}{\sqrt{-g}}
\partial_{\beta}\left(\sqrt{-g}F^{\alpha\beta}\right)
=0$ has as 
a constant solution for the inner 
electric potential $\phi_i$
which, for convenience, can be written as
\begin{equation}\label{phiin}
\phi_i= \frac{Q}{R}+{\rm constant}\,,\quad r< R\,,
\end{equation}
where $Q$ is a constant, 
to be interpreted as the conserved electric charge.
In the outer region $\mathcal{V}_o$ ($r> R$), 
the spacetime is Reissner-Nordstr\"om
and the 
Maxwell equation 
$\nabla_{\beta}F^{\alpha\beta}=\frac{1}{\sqrt{-g}}
\partial_{\beta}\left(\sqrt{-g}F^{\alpha\beta}\right)
=0$ now yields
\begin{equation}\label{phiout}
\phi_o= \frac{Q}{r}+{\rm constant}\,,\quad r>R\,.
\end{equation}

Due to the existence of electricity in the shell, another important
set of restrictions must also be considered.  These restrictions are
related to the discontinuity present in the electric field across the
charged shell. We are interested in the projection
\begin{equation}
A_a = A_{\al} \,e^{\al}_a
\end{equation}
of the four-potential in the shell's
hypersurface, since it will contain
quantities which are intrinsic to the shell. 
Indeed, following \cite{Kuchar},
\begin{equation}\label{potjunct1}
\left[A_{a}\right]=0\,,
\end{equation}
with
$A_{i\,a} = (-\phi_{i}, 0,0)$, and 
$A_{o\,a} = (-\phi_{o}, 0,0)$
being the vector potential at $R$, 
on the shell, seen from each side of it.  
Thus, the constants in Eqs.~(\ref{phiin}) and
(\ref{phiout}) are indeed the same and so
at $R$
\begin{equation}
\phi_o =
\phi_i \,,\quad r=R\,.
\end{equation}
Following \cite{Kuchar} further, the tangential
components $F_{ab}$ of the electromagnetic tensor $F_{\al\be}$ must
change smoothly across $\Sigma$, i.e.
\begin{equation}\label{JC1}
\left[F_{ab}\right]=0\,,
\end{equation}
with
\begin{equation}
F^{i}_{ab} = F^{i}_{\al\be} e^{\al}_{i}{}_a \,
e^{\be}_{i}{}_b\,,\quad
F^{o}_{ab} = F^{o}_{\al\be} e^{\al}_{o}{}_a \,
e^{\be}_{o}{}_b\,,
\end{equation}
while the normal components $F_{a\perp}$ must change by a jump
as,
\begin{equation}\label{JC2}
\left[F_{a\perp}\right] = 4\pi \sigma_e u_{a}\,,
\end{equation}
where 
\begin{equation}
F^{i}_{a\perp} = F^{i}_{\al\be}  e^{\al}_{i}{}_a \, 
n^{\be}_{i}\,,\quad
F^{o}_{a\perp} = F^{o}_{\al\be}  e^{\al}_{o}{}_a \, 
n^{\be}_{o}\,,
\end{equation}
and $\sigma_eu_{a}$ is the surface electric current,
with  $\sigma_e$ being the density of charge and
$u_{a}$ its 3-velocity, defined on the shell.
One can then show that Eq.~(\ref{JC1}) is
trivially satisfied, while Eq.~(\ref{JC2}) leads to the single
nontrivial equation at $R$, on the shell, 
\begin{equation}\label{JCn1}
\frac{\partial\phi_o}{\partial r}
-\frac{\partial\phi_i}{\partial r} 
= - 4\pi \sigma_e\,,\quad r=R\,.
\end{equation}
Then, from Eqs.~(\ref{phiin}), (\ref{phiout}),
and (\ref{JCn1}) one obtains
\begin{equation}\label{PhiJC}
\frac{Q}{R^2}=4\pi \sigma_e \,,
\end{equation}
relating the total charge $Q$, the charge density $\sigma_e$,
and the shell's radius $R$ in the expected manner.
This section with its equations 
forms the dynamical side of the electric thin shell
solution.

\subsection{Restrictions on the thin-shell radius}
\label{restric}

A natural inequality that the
shell should obey is to consider the shell to be outside its
gravitational radius in all instances, so
\begin{equation}
R \geq r_+.
\label{notrapped}
\end{equation}
It is then clear 
that the
physical allowed values for $k$ in Eq.~(\ref{red})
are in the interval $[0,1]$.
It is also interesting to consider the restrictions 
imposed by the dominant energy condition.
However, since it will not take part in our analysis
we leave this discussion for Appendix \ref{apa}.

\section{Thermodynamics
and stability conditions for the thin shell: Generics}
\label{thermo}

\subsection{Thermodynamics and integrability
conditions for the thin shell}
\label{thermo2}

We now turn to the thermodynamic 
side and to the
calculation of the entropy of the shell. 
We use units in which the Boltzmann constant is one.
We start
with the assumption that the shell 
in static equilibrium
possesses a well-defined
temperature $T$ and an entropy $S$ which is a function of 
three variables, call them $M$, $A$, $Q$, i.e.,
\begin{equation}\label{entropy0}
S=S(M,A,Q)\,.
\end{equation}
$(M,A,Q)$ can be considered as three generic parameters.
In our connection they are
the shell's
rest mass $M$, area $A$, and charge $Q$.
The first law of thermodynamics can thus be written as
\begin{equation}\label{TQ}
T dS = dM + pdA - \Phi dQ
\end{equation}
where $dS$
is the differential of the entropy of the shell,
$dM$ is the
differential of the rest mass,
$dA$ is the differential of the area of the shell,
$dQ$ is the differential of the charge,
and $T$, $p$ and
$\Phi$ are the temperature, the pressure, and the thermodynamic 
electric potential of the shell, respectively. 
In order to find the entropy $S$,
one thus needs three equations of state, namely,
\begin{equation}\label{press0}
p=p(M,A,Q)\,,
\end{equation}
\begin{equation}\label{temper}
\beta=\beta(M,A,Q)\,,
\end{equation}
\begin{equation}\label{electr}
\Phi=\Phi(M,A,Q)\,,
\end{equation}
where
\begin{equation}\label{beta}
\be \equiv \frac1T
\end{equation}
is the inverse temperature.

It is important to note that the temperature 
and the thermodynamic
electric potential
play the role of
integration factors, which implies that there will be integrability
conditions that must be specified in order to guarantee the existence
of an expression for the entropy, i.e. that the differential $dS$ is
exact. These integrability conditions are
\begin{align}
\left(\frac{\dd \be}{\dd A}\right)_{M,Q} & = \hspace{3mm}
\left(\frac{\dd \be p}{\dd M}\right)_{A,Q}\,, \label{Ione} \\
\left(\frac{\dd \be}{\dd Q}\right)_{M,A} & = - \left(\frac{\dd \be
\Phi}{\dd M}\right)_{A,Q}\,, \label{Itwo} \\
\left(\frac{\dd \be p}{\dd Q}\right)_{M,A} & = - \left(\frac{\dd \be
\Phi}{\dd A}\right)_{M,Q} \,.
\label{Ithree}
\end{align}
These equations enable one to determine the 
relations between the three equations of state of
the system.

\subsection{Stability conditions for the thin shell}
\label{thermo3}

With the first law of thermodynamics given in Eq.~(\ref{TQ}),
one is able
to perform a thermodynamic study of the local
intrinsic stability of the shell.
To have thermodynamic stability the
following inequalities should hold
\begin{equation}\label{B1}
\left(\frac{\dd^2 S}{\dd M^2}\right)_{A,Q} \leq 0\,,
\end{equation}
\begin{equation}\label{B2}
\left(\frac{\dd^2 S}{\dd A^2}\right)_{M,Q} \leq 0\,,
\end{equation}
\begin{equation}\label{B3}
\left(\frac{\dd^2 S}{\dd Q^2}\right)_{M,A} \leq 0\,,
\end{equation}
\begin{equation}\label{B4}
\left(\frac{\dd^2 S}{\dd M^2}\right)\left(\frac{\dd^2 S}{\dd
A^2}\right) - \left(\frac{\dd^2 S}{\dd M \dd A}\right)^2 \geq 0\,,
\end{equation} 
\vspace{2mm}
\begin{equation}\label{B5}
\left(\frac{\dd^2 S}{\dd A^2}\right)\left(\frac{\dd^2 S}{\dd
Q^2}\right) - \left(\frac{\dd^2 S}{\dd A \dd Q}\right)^2 \geq 0\,,
\end{equation}
\vspace{2mm}
\begin{equation}\label{B6}
\left(\frac{\dd^2 S}{\dd M^2}\right)\left(\frac{\dd^2 S}{\dd
Q^2}\right) - \left(\frac{\dd^2 S}{\dd M \dd Q}\right)^2 \geq 0\,,
\end{equation} 
\vspace{2mm}
\begin{equation}\label{B7}
\left(\frac{\dd^2 S}{\dd M^2}\right) \left(\frac{\dd^2 S}{\dd Q \dd
A}\right) - \left(\frac{\dd^2 S}{\dd M \dd A}\right) \left(\frac{\dd^2
S}{\dd M \dd Q}\right) \geq 0\,.
\end{equation}
The derivation of these expressions 
follows the rationale presented in
\cite{callen}, see Appendix 
\ref{apb}.

\section{The thermodynamic independent variables
and the three equations of state: equations for the pressure,
temperature and electric potential}
\label{eqsos}

\subsection{The three independent 
thermodynamic variables $(M,R,Q)$}

We will work from now onwards with the three 
independent variables $(M,R,Q)$ instead of
$(M,A,Q)$.
The rest mass $M$ of the shell is from
Eq.~(\ref{sigmarestmasssigma}) given by
\begin{equation}
M = 4\pi R^2 \, \sigma \,,
\label{restmasssigma}
\end{equation}
where $\sigma$ is given by Eq.~(\ref{Sig1})
and $R$ is the radius of the shell.
The first law of thermodynamics 
written in generic terms 
is simpler when expressed
using the area $A$ of the shell, but here
it is 
handier to use the
radius $R$ in this specific study. 
The radius $R$ is related to the area $A$
through Eq.~(\ref{intrinsmetr}), i.e., 
\begin{equation}
R= \sqrt
{
\dfrac{A}{4\,\pi}}.
\label{radiusarea}
\end{equation}
As for the charge $Q$,
using Eq.~(\ref{PhiJC}), it is given by
\begin{equation}
Q =  4\pi R^2 \, \sigma_e .
\end{equation}
The three independent 
thermodynamic variables are thus $(M,R,Q)$.

We should now envisage 
Eq.~(\ref{m0}) and 
Eqs.~(\ref{horradi})-(\ref{horradicauch})
as functions of $(M,R,Q)$, i.e.
\begin{equation}\label{m}
m(M,R,Q) = M - \frac{G M^2}{2R} + \frac{Q^2}{2R}\,,
\end{equation}
and
\begin{equation}
r_{+}(M,R,Q) = G\,m(M,R,Q) +\sqrt{G^2m(M,R,Q)^2 - G Q^2}\,,
\label{horradi2}
\end{equation}
\begin{equation}
r_{-}(M,R,Q) = G\,m(M,R,Q) - \sqrt{G^2m(M,R,Q)^2 - G Q^2}\,,
\label{horradicauch2}
\end{equation}
respectively.
The function $k$ in Eq.~(\ref{red}) is also a function of 
$(M,R,Q)$,
\begin{align}\label{red2}
&k(r_{+}(M,R,Q),r_{-}(M,R,Q),R)=&\nonumber\\
&\sqrt{\Big(1-\frac{r_+(M,R,Q)}{R}\Big)\Big(1-
\frac{r_-(M,R,Q)}{R}\Big)}\,.
\end{align}

\subsection{The pressure equation of state}

Expressing the pressure equation of state in the form of
Eq.~(\ref{press0}), we obtain from Eqs.~(\ref{SS2})
and (\ref{m0}) [or Eq.~(\ref{m})],
\begin{equation}\label{pinMRQ}
p(M,R,Q) = \frac{GM^2-Q^2}{16 \pi R^2 (R-GM)}\,, 
\end{equation}
or changing from the variables $(M,R,Q)$ to $(r_+,r_-,R)$
which is more useful, we find
[see Eqs.~(\ref{pQk1}) and (\ref{invhorradicauch})],
\begin{align} \label{pQk}
&p(r_+,r_-,R) =  \nonumber \\
&\frac{R^2(1-k)^2
 - r_+ r_-}{16 \pi G R^3 \,k}\,, 
\end{align}
where  $k$ can be envisaged as $k=k(r_+,r_-,R)$ 
as given in Eq.~(\ref{red2})
and $r_+$ and $r_-$ are functions of $\,(M,R,Q)$, see 
Eqs.~(\ref{horradi2})-(\ref{horradicauch2}).
This 
reduces to the expression obtained in \cite{Mart} in the limit
$Q=0$ or $r_-=0$. 
This equation, Eq.~(\ref{pQk}), is a pure 
consequence of the Einstein equation, encoded in the junction
conditions.

\subsection{The temperature equation of state}

Turning now to the temperature equation of state (\ref{temper}), we
will need to focus on the integrability condition
(\ref{Ione}). Changing 
from the variables $(M,R,Q)$ to $(r_+,r_-,R)$, Eq.~(\ref{Ione})
becomes
\begin{equation}
\left(\frac{\dd \be}{\dd R}\right)_{r_+,r_-} = 
\be \frac{R(r_+ + r_-)-2r_+ r_-}{2 R^3 k^2}
\end{equation}
which has the analytic solution
\begin{equation}\label{BS5}
\be(r_+,r_-,R) = b(r_+,r_-) \,k
\end{equation}
where 
$k$ is 
a function of $r_+$, $r_-$, and $R$, as given in 
Eq.~(\ref{red2}), and 
$b(r_+,r_-)\equiv \be(r_+,r_-,\infty)$ is an arbitrary function,
representing the inverse of the temperature of the shell if its radius
were infinite.
Hence, in a sense, from Eq.~(\ref{BS5}),
we recover Tolman's formula for the temperature of
a body in curved spacetime. The arbitrariness of this
function is due to the fact that the matter fields of the  shell
are not specified. Note that $b$ and $k$ 
are still functions of $(M,R,Q)$
through the variables $r_+$ and $r_-$, see 
Eqs.~(\ref{horradi2})-(\ref{horradicauch2})
and Eq.~(\ref{red2}).

\subsection{The electric potential equation of state}

The remaining equation of state to be studied is the electric
potential.  Using Eqs.~(\ref{M1}) and (\ref{red2}), one can deduce
$\left(\frac{\partial M}{\partial A}\right)_{r_{+},r_{-}}=-p$, i.e.,
\begin{equation}
\left(\frac{\partial M}{\partial R}\right)_{r_{+},r_{-}}=-8\pi\,R\,p\,. 
\label{mp}
\end{equation}
Then, it follows from Eqs.~(\ref{Ione})-(\ref{Ithree}) 
and Eq.~(\ref{mp}) that the differential
equation
\begin{equation}
\left(\frac{\dd p}{\dd Q}\right)_{M,R} + \frac{1}{8 \pi
R}\left(\frac{\dd \Phi}{\dd R}\right)_{r_+,r_-} + \Phi \left(\frac{\dd
p}{\dd M}\right)_{R,Q} = 0\,,
\label{phieqdiff}
\end{equation}
holds, where the second term has been expressed in the variables
$(r_+,r_-,R)$ and the other terms in the variables $(M,R,Q)$ for the
sake of computational simplicity. 
Then, after using Eq.~(\ref{pinMRQ})
in Eq.~(\ref{phieqdiff}), 
we obtain that Eq.~(\ref{phieqdiff}) takes the form
\begin{equation}
R^{2}\,
\left(\frac{\partial \Phi k}{\partial R}\right)_{r_+,r_-}
- \frac{\sqrt{r_+ r_-}}
{\sqrt{G}}=0\,,
\label{above}
\end{equation}
where $k$ can be envisaged as $k=k(r_+,r_-,R)$ as given in
Eq.~(\ref{red2}). The solution of Eq.~(\ref{above}) is then
\begin{equation}\label{Phi0}
\Phi(r_+,r_-,R) = \frac{\phi(r_+,r_-) - \frac{\sqrt{r_+ r_-}}
{\sqrt{G}R}}{k}
\end{equation} 
where $\phi(r_+,r_-)\equiv \Phi(r_+,r_-,\infty)$ is an arbitrary
function that corresponds physically to the electric potential of the
shell if it were located at infinity.  This thermodynamic electric
potential $\Phi$ is the difference in the 
electric potential $\phi$ between
infinity and $R$, blueshifted from infinity to $R$ (see a similar
result in \cite{yorketal,pecalemos} for an
electrically charged black hole in a grand canonical
ensemble).  We also see that, once again, by changing to the variables
$(r_+,r_-,R)$ we are able somehow
to reduce the number of arguments of the
arbitrary function from three to two.

It is convenient to define a function $c(r_+,r_-)$
through $c(r_+,r_-) \equiv \frac{\phi(r_+,r_-)}{Q}$, i.e.,
\begin{equation}\label{cr+r-}
c(r_+,r_-) \equiv \sqrt{G} \frac{\phi(r_+,r_-)}{\sqrt{r_+ r_-}}\,,
\end{equation} 
where we have used $Q=\sqrt{r_+r_-/G}$
as given in Eq.~(\ref{invhorradicauch}).
Then, Eq.~(\ref{Phi0}) is written as
\begin{equation}\label{Phi}
\Phi(r_+,r_-,R) = \dfrac{ {c(r_+,r_-)}
- \dfrac{1}{R}  }
{k}\,\sqrt{\frac{r_+r_-}{G}}\,,
\end{equation} 
where  $k$ can be envisaged as $k=k(r_+,r_-,R)$ 
as given in Eq.~(\ref{red2}). 

\section{Entropy of the thin shell}
\label{entro}

At this point we have all the necessary information to calculate the
entropy $S$. By inserting the 
equations of state for the pressure, 
Eq.~(\ref{pQk}),
for the temperature, 
Eq.~(\ref{BS5}), and for the
electric potential, 
Eq.~(\ref{Phi}), as well as 
the differential of $M$ given in
Eq.~(\ref{M1}) and the differential of
the area $A$ or of the radius $R$, see 
Eq.~(\ref{radiusarea}),
into the first law,
Eq.~(\ref{TQ}),
we
arrive at the entropy differential
\begin{align}\label{dSQ}
dS = b(r_+,r_-) & \frac{1-c(r_+,r_-) r_-}{2G} dr_+ \nonumber \\ 
     & + b(r_+,r_-)\frac{1-c(r_+,r_-) r_+}{2G} dr_-\,,
\end{align}
Now, Eq.~(\ref{dSQ}) has its own integrability condition if $dS$ is to
be an exact differential. Indeed, it must satisfy the equation
\begin{equation}\label{bc5}
\frac{\dd b}{\dd r_-} (1-r_- c) - \frac{\dd b}{\dd r_+}(1-r_+ c) =
\frac{\dd c}{\dd r_-} b r_- - \frac{\dd c}{\dd r_+} b r_+.
\end{equation}
This shows that in order to obtain a specific expression for the
entropy one can choose either $b$ or $c$,
and the other remaining function can
be obtained by solving the differential equation (\ref{bc5}) with
respect to that function. Since Eq.~(\ref{bc5}) is 
a differential equation there is still some freedom
in choosing the other remaining function. 
In the first examples 
we will choose to specify the
function $b$ first and from it obtain an expression for $c$.
We also give examples where 
the function $c$ is specified first.

From Eq.~(\ref{dSQ}) we obtain
\begin{equation}\label{entr5}
S=S(r_+,r_-)\,,
\end{equation}
so that the entropy is a function of $r_+$ and $r_-$ alone.
In fact $S$ is a function of $(M,R,Q)$, 
$S(M,R,Q)$,
but the functional dependence has to be 
through $r_+(M,R,Q)$ and $r_-(M,R,Q)$,
i.e., in full form
\begin{equation}\label{entr6}
S(M,R,Q)=S(r_+(M,R,Q),r_-(M,R,Q))\,.
\end{equation}
This result shows that the entropy of the thin 
charged shell
depends on the $(M,R,Q)$ through 
$r_+$ and $r_-$ which themselves
are specific functions of $(M,R,Q)$.

It is also worth noting the following feature.
From Eq.~(\ref{entr6}) we see that 
shells with the same $r_+$ and $r_-$, i.e., the 
same ADM mass $m$ and charge $Q$, 
but different radii $R$, have the
same entropy. 
Let then 
an observer sit at infinity and measure $m$
and $Q$ (and thus $r_+$ and $r_-$). 
Then, the observer
cannot distinguish the entropy of shells with
different radii.  This is a kind of thermodynamic
mimicker, as a shell near its own 
gravitational radius and another one far from
it have the same entropy.

\section{The thin shell and the black hole limit}
\label{bhlimit}

\subsection{The temperature equation of state
and the entropy}
\label{peos}

Let us consider a charged thin shell, for which the differential of
the entropy has been deduced to be Eq.~(\ref{dSQ}).
We are free 
to choose an equation of state for the inverse temperature.
Let us pick for convenience the following inverse temperature
dependence,
\begin{equation}\label{bBH}
b(r_+,r_-) = \gamma\,
\frac{r_+^2}{r_+-r_-}\,,
\end{equation}
where $\gamma$ is some constant with 
units of inverse mass times inverse radius,
i.e., units of angular momentum.

For a charged shell we must also specify the function $c(r_+,r_-)$,
whose form can be taken from 
the differential equation (\ref{bc5}) upon
substitution of the function (\ref{bBH}). 
There is a family of solutions 
for $c(r_+,r_-)$
but for
our purposes here we choose 
the following specific solution,
\begin{equation}\label{cBH}
c(r_+,r_-) = \frac{1}{r_+}\,.
\end{equation}
The rationale for the choices above becomes clear when
we discuss the shell's gravitational radius, i.e., black hole,
limit.
Inserting the choice for $b(r_+,r_-)$, Eq.~(\ref{bBH}),
along with the  choice for the function
$c(r_+,r_-)$,  Eq.~(\ref{cBH}), in the
differential (\ref{dSQ}) and integrating, we obtain the entropy 
differential for the
shell
\begin{equation}\label{diffSS}
dS = \frac{\gamma}{2\, G} \,r_+\, dr_+\,.
\end{equation}
Thus, the entropy of the shell is 
$S=\frac{\gamma}{4\, G} \,r_+^2+ S_0$,
where $S_0$ is an integration constant.
Imposing that when the shell vanishes 
(i.e., $M=0$ and $Q=0$, and so 
$r_+=0$) the entropy vanishes 
we have that $S_0$ is zero, 
and so 
$S=\frac{\gamma}{4\, G} \,r_+^2$.
Thus, we can write
the entropy $S(M,R,Q)$ as 
\begin{equation}\label{SSa}
S = \frac{\gamma}{16\pi G}\, A_+\,,
\end{equation}
where $A_+$ is the gravitational 
area of the shell, as given in 
Eq.~(\ref{arear+}).
This result shows that the entropy of this thin 
charged shell
depends on $(M,R,Q)$ through 
$r_+^2$ only, which itself is
a specific function of $(M,R,Q)$.

Now, what is the constant $\gamma$? It should be determined
by the properties of the matter in the shell,
and cannot be decided a priori.

\subsection{The stability conditions for the specific temperature
ansatz}
\label{stabil1}

The thermodynamic stability of the uncharged case ($Q=0$, i.e.,
$r_-=0$) can be worked out
\cite{Mart} and elucidates the issue.  In the
uncharged case the nontrivial stability conditions are given by
Eqs.~(\ref{B1}) and (\ref{B4}). 
Equation~(\ref{B1}) gives immediately
$R\leq\frac32 r_+$, i.e., $R\leq3Gm$. On the other hand,
Eq.~(\ref{B4}) yields $R\geq r_+$, i.e., $R\geq2Gm$. Thus, the
stability conditions yield the following range for $R$, $r_+\leq
R\leq\frac32 r_+$, or in terms of $m$, $2Gm\leq R\leq 3Gm$. This is
precisely the range for stability found by York \cite{york1} for a
black hole in a canonical ensemble in which a spherical massless thin
wall at radius $R$ is maintained at fixed temperature $T$. In
\cite{york1} the criterion used for stability is that the heat
capacity of the system should be positive, and physically such a tight
range for $R$ means that only when the shell, at a given temperature
$T$, is sufficiently close to the horizon can it smother the
black hole enough 
to make it thermodynamically stable. The positivity of the heat
capacity is equivalent to our stability conditions, Eqs.~(\ref{B1})
and (\ref{B4}) in the uncharged case.

The stability conditions, Eqs.~(\ref{B1})-(\ref{B7}), for the general
charged case cannot be solved analytically in this instance, they
require numerical work, which will shadow what we want to
determine. Nevertheless, the approach followed in
\cite{yorketal,pecalemos} for the heat capacity of a charged black
hole in a grand canonical ensemble gives a hint of the procedure that
should be followed.

\subsection{The black hole limit}
\label{bhl}

\subsubsection{The black hole limit properly stated}
\label{bhlps}

Although $\gamma$ should be determined
by the properties of the matter in the shell,
there is a case in which the properties
of the shell have to adjust to the 
environmental properties of the spacetime. 
This is the case when
$R
\to r_+$.
In this case, one must note that, as the shell approaches its own
gravitational radius, quantum fields are inevitably present and their
backreaction will diverge unless we choose the 
black hole Hawking temperature
$T_{\rm bh}$
for the temperature of the shell.
In this case, $R\to r_+$, 
we must take the temperature of 
the shell as
$T_{\rm bh} = \frac{\hbar}{4 \pi} \frac{r_+-r_-}{r_+^2}$,
where $\hbar$ is Planck's constant.
So we must choose
\begin{equation}
\gamma= \frac{4\pi}{\hbar} \,,
\label{ga}
\end{equation}
i.e., $\gamma$ depends on fundamental constants.
Then,
\begin{equation}\label{thawk}
b(r_+,r_-) = \frac{1}{T_{\rm bh}} 
=\frac{4 \pi} {\hbar}
\frac{r_+^2}{r_+-r_-}\,.
\end{equation}
In this case the entropy of the shell is 
$S= \frac14\frac{A_+}{G\hbar}$, i.e., 
\begin{equation}\label{SS}
S= 
\frac{1}{4}\frac{A_+}{A_p}\,,
\end{equation}
where $l_p=\sqrt{G\hbar}$ is the Planck length, and $A_p=l_p^2$ the
Planck area.  
Note now that the entropy
given in Eq.~(\ref{SS})
is the black hole Bekenstein-Hawking entropy
$S_{\rm bh}$ of a charged black hole since
\begin{equation}\label{SSBH} 
S_{\rm bh}= \frac{1}{4}\frac{A_+}{A_p}\,,
\end{equation} 
where $A_+$ is here the horizon area.  Thus, when we
take the shell to its own gravitational radius the entropy is the
Bekenstein-Hawking entropy.  The limit also implies that the pressure
and the thermodynamic electric potential go to infinity as $1/k$,
according to Eqs.~(\ref{pQk}) and (\ref{Phi}), respectively.  Note,
however, that the local inverse temperature goes to zero as $k$, see
Eq.~(\ref{BS5}), and so the local temperature of the shell also goes
to infinity as $1/k$. These well-controlled infinities cancel out
precisely to give the Bekenstein-Hawking entropy (\ref{SS}).

Note that, since $A=A_+$ when the shell is at its own gravitational
radius, at this point the entropy of the shell is proportional to its
own area $A$, indicating that all the shell's fundamental degrees of
freedom have been excited.

Note also that the shell at its own gravitational radius, at least in
the uncharged case, is thermodynamically stable, since in this case
stability requires $r_+\leq R\leq\frac32 r_+$, as mentioned above.

Note yet that our approach and the approach followed in
\cite{pretvolisr} to find the black hole entropy have some
similarities. The two approaches use matter fields, i.e., shells,
to find the black hole entropy. Here we use a static shell
that decreases its own radius $R$ by steps, maintaining its staticity
at each step. In \cite{pretvolisr} a reversible contraction of a thin
spherical shell down to its own gravitational radius was examined, and
it was found that the black hole entropy can be defined as the
thermodynamic entropy stored in the matter in the situation that the
matter is compressed into a thin layer at its own gravitational
radius.

Finally we note that the extremal limit $\sqrt{G}m=Q$ or $r_+=r_-$
is well defined from above.  Indeed, when one takes the limit $r_+\to
r_-$ one finds that $1/b(r_+,r_-)=0$ (i.e., the Hawking temperature is
zero) and the entropy of the extremal black hole is still given by
$S_{\rm extremal\,bh}= \frac{1}{4}\frac{A_+}{A_p}$.  It is well known
that extremal black holes and in particular their entropy have to be
dealt with care.  If, ab initio, one starts with the analysis for an
extremal black hole one finds that the entropy of the extremal black
hole has a more general expression than simply being equal to one
quarter of the area \cite{pretvolisr,lemoszaslaextremal}.  This
extremal shell is an example of a Majumdar-Papapetrou matter system.
Its pressure $p$ is zero, and it remains zero, and thus finite, even
when $R\to r_+$.  This limit of $R\to r_+$ is called a quasiblack
hole, which in the extremal case is a well-behaved one.

\subsubsection{The rationale for the choice of 
$b(r_{+},r_{-})$ and $c(r_{+},r_{-})$}
\label{rational}

We have started with a thin shell and imposed a temperature equation
of state of the Hawking type, see Eq.~(\ref{bBH}) [see also
Eqs.~(\ref{ga}), and (\ref{thawk})], and a specific thermodynamic
electric potential, see Eq.~(\ref{cBH}).  This set of equations of
state gives an entropy for the shell proportional to its gravitational
radius area $A_+$.  One can moreover set the temperature of the shell
at any $R$ precisely equal to the Hawking temperature
$T_{\mathrm{bh}}$, see Eq.~(\ref{thawk}). Remarkably, we have then
shown that a self-gravitating electric thin shell at the Hawking
temperature and with a specific electric potential has a
Bekenstein-Hawking entropy.

A priori, Hawking-type choices for the temperature 
[Eqs.~(\ref{bBH}), (\ref{ga}), and (\ref{thawk})],
and black hole-type choices for the electric potential 
[Eq.~(\ref{cBH})], are simply choices,
many other choices for the set of equations of state can be
taken. However, this set is really imposed on the shell when it
approaches its gravitational radius, where it takes the precise forms
given in Eqs.~(\ref{bBH}) and (\ref{ga})
[or, Eq.~(\ref{thawk})],
and (\ref{cBH})
as the spacetime quantum effects get a hold
on the shell.

We would like to stress that the requirement $b=T_{\mathrm{bh}}^{-1}$
[see Eqs.~(\ref{bBH}), (\ref{ga}), and (\ref{thawk})] is compulsory
only for shells that approach their own gravitational
radius. Otherwise, if we consider the radius of the shell within some
constrained region outside the gravitational radius, the shell
temperature can be arbitrary since away from the horizon, quantum
backreaction remains modest and does not destroy the thermodynamic
state. 
One can discuss whole classes of functions $b(r_{+},r_{-})\neq
T_{\mathrm{bh}}^{-1}$.

In addition we stress 
that the choice for $c$, Eq.~(\ref{cBH}), is necessary only
for shells at the gravitational radius limit.
According to Eq.~(\ref{cr+r-}), this gives us 
$\phi =\frac{ \sqrt{r_-} }{ \sqrt{G\, r_{+} }}$, i.e., 
\begin{equation}
\phi =\frac{Q}{r_{+}}  \label{rn}
\end{equation}
that coincides with the standard expression for the electric potential
for the Reissner-Nordstr\"{o}m black hole. In addition, 
Eq.~(\ref{Phi})
acquires the form
\begin{equation}
\Phi =\frac{Q}{k}\left(\frac{1}{r_{+}}-\frac{1}{R}\right) 
\label{frn}
\end{equation}
that coincides entirely with the corresponding formula for the
Reissner-Nordstr\"{o}m black hole in
a grand canonical ensemble \cite{yorketal}.  Meanwhile, in our
case there is no black hole.  Moreover, if we go to the uncharged
case, $Q\rightarrow0$ or 
$r_{-}\rightarrow 0$, and thus the outer space is described by the
Schwarzschild metric, then it is seen from Eq.~(\ref{dSQ}) that the
quantity $c$ drops out from the entropy, so the choice of $c$ is
relevant for the charged case only, of course.

\subsubsection{Similarities between the thin-shell approach and
the black hole mechanics approach}
\label{sim}

There are similarities between the thin-shell approach and
the black hole mechanics approach \cite{hawk1}.
These are evident if we express the
differential of the entropy of the charged shell (\ref{dSQ}) in terms
of the black hole ADM 
mass $m$ and charge $Q$, given in terms of the
variables $(r_+,r_-)$ by 
Eqs.~(\ref{invhorradi})-(\ref{invhorradicauch}).
The differential for the entropy of the shell reads
in these variables
\begin{equation}
T_0 dS = dm - c\, Q\, dQ
\end{equation}
where we have defined $T_0 \equiv 1/b(r_+,r_-)$ which is the
temperature the shell would possess if located at infinity.
Here, $T_0 =1/b(r_+,r_-)$ and $c=c(r_+,r_-)$ should be seen as 
$T_0 (m,Q)=1/b(m,Q)$ and $c(m,Q)$, respectively,
since $r_+$ and $r_-$ are functions of
$m$ and $Q$.
As we have
seen, if we take the shell to its gravitational radius, we must fix
$T_0 = T_{\rm bh}$ and $c = 1/r_+$.  
This suggests that $Q/r_+$ should play the role of
the black hole electric potential $\Phi_{\rm bh}$, which in fact is
true, as shown in Eq.~(\ref{rn}) 
(see \cite{hawk1}, see also \cite{yorketal,pecalemos}).
So the
conservation of energy of the shell is expressed as
\begin{equation}\label{1T7}
T_{\rm bh} dS_{\rm bh} = dm - \Phi_{\rm bh}\, dQ\,.
\end{equation}
We thus
see that the first law of thermodynamics for the shell 
at its own gravitational radius is 
equal to the energy conservation for the 
Reissner-Nordstr\"om
black hole.

\section{The thin shell
with another specific equation of state for the 
temperature}
\label{eqstate}

\subsection{The temperature equation of state
and the entropy}

The previous equation of state is not prone to a simple stability 
analysis. Here we give another equation of state 
that permits finding both  an expression for the shell's entropy
and performing a simple stability analysis.

We must first
specify an adequate thermal equation of state for $b(r_+,r_-)$. 
A possible
simple choice is a power law in the ADM mass $m$, i.e., 
$b(r_+,r_-)$ has the form
\begin{equation}\label{bF6}
b(r_+,r_-) = 2G \, a(r_++r_-)^{\al}
\end{equation}
where $a$ and $\al$ are free coefficients
related to the properties of the shell. 
Power laws occur frequently in thermodynamic systems, 
and so this is a natural choice as well.
The simple choice above 
allows one to find the form of the function $c$. 
Indeed, the integrability equation
(\ref{bc5}) gives that the function $c$ 
can be put in the form
$c(r_+,r_-) = 2G \frac{f(r_+ r_-)}{(r_+ + r_-)^{\al}}$, where
$f(r_+ r_-)$ is an arbitrary function
of the product $r_+ r_-$ and supposedly also depends on the
intrinsic constants of the matter that makes up the shell.
For convenience we choose $f(r_+ r_-)=d\,(r_+ r_-)^{\delta}$,
where $d$ and $\delta$ are parameters
that reflect the shell's properties, so that
\begin{equation}\label{cpower}
c(r_+,r_-) = 2G \,d\,  \frac{(r_+ r_-)^{\delta}}
{(r_+ + r_-)^{\al}}\,.
\end{equation}
The gravitational constant $G$ was introduced in 
Eqs.~(\ref{bF6}) and (\ref{cpower}) for convenience. 
Inserting Eqs.~(\ref{bF6})-(\ref{cpower})
into Eq.~(\ref{dSQ}) and integrating, 
gives the entropy
\begin{equation}\label{SF6}
S(r_+,r_-) = a \, \left[\frac{(r_++r_-)^{\al+1}}{\al+1} - d \,
\frac{(r_+ r_-)^{\de+1}}{\de+1}\right] \,,
\end{equation}
where the constant of integration $S_0$ 
has been put to zero, as expected in the limit $r_+
\to 0$ and $r_- \to 0$.
Again, the entropy of this thin 
charged shell
depends on $(M,R,Q)$ through 
$r_+$ and $r_-$ only, which in turn are
specific functions of $(M,R,Q)$.

We consider positive temperatures and 
positive electric potentials, so 
\begin{equation}\label{adgeq0}
a>0\,,\quad d>0\,.
\end{equation}
We consider only 
\begin{equation}\label{ageq0}
\al>0 \,,
\end{equation}
for the simplicity of the upcoming stability analysis. Although this
choice somewhat narrows down the range of cases to which the analysis
is applicable, it only rules out the cases where $-1<\al<0$, since for
values $\al\leq-1$ it would give a diverging entropy in the limit $r_+
\to 0$ and $r_- \to 0$, something which is not physically
acceptable. Indeed, in such a limit we would expect the entropy to be
zero which requires $\al > -1$.

\subsection{The stability conditions for the specific temperature
ansatz}
\label{stabil2}

Proceeding to the thermodynamic stability treatment, we start with
Eq.~(\ref{B1}), which can be shown to be equivalent to
\begin{equation}\label{B1s}
r_+ r_- - 2 R^2 k^2 \al + (1-k^2)R^2 \geq 0.
\end{equation}
Solving for $k$, this leads to the restriction
\begin{equation}\label{k1s}
k \leq \sqrt{\frac{1}{2\al +1}\left(1+\frac{r_+ r_-}{R^2}\right)}.
\end{equation}

Going now to Eq.~(\ref{B2}), it gives
\begin{align}\label{B2s}
[r_+ r_- - (1-k)^2 R^2]&[\al (r_+ r_- - (1-k)^2 R^2) \nonumber \\
& + 3(r_+ r_- + (1-k^2)R^2)] \leq 0.
\end{align}
Since the second multiplicative term on the left must be positive, 
one can solve for $k$ and obtain the set of values which satisfy 
the inequality,
\begin{align}\label{k2s}
\frac{\al}{\al+3} - \sqrt{\frac{9}{(\al+3)^2} + \frac{r_+ r_-}{R^2}} 
\leq k  \nonumber \\
\leq\frac{\al}{\al+3} + \sqrt{\frac{9}{(\al+3)^2} + 
\frac{r_+ r_-}{R^2}}\,.
\end{align}
As for Eq.~(\ref{B3}), it reduces to
\begin{equation}
\frac{d R(2\de +1)(r_+ r_-)^{\de}}{\left(\frac{r_+ r_-}{R} + 
(1-k^2)R\right)^{\al}} \geq \frac{R^2(1-k^2)+(2 \al + 1)r_
+ r_-}{R^2(1-k^2)+r_+ r_-}.
\end{equation}
Although one cannot conclude anything directly from the above 
inequality, it is nonetheless worth noting that the right-hand
side is greater than zero, and so $\de$ must obey the condition
\begin{equation}\label{intDe}
\de \geq -\frac{1}{2}.
\end{equation}

Regarding Eq.~(\ref{B4}), it is possible to show that it implies 
the condition
\begin{align}\label{B4s}
r_+^2 r_-^2 & (\al+3) - 2r_+ r_-R^2(2k^2\al + 2k^2 - 
k + \al -1) \nonumber \\
& + (1-k)^2 R^4 (3k^2 \al + k^2 + 2\al k + \al -1) \leq 0,
\end{align}
which does not provide any information on its own since it is 
a polynomial of order four in the variable $k$. Nonetheless, 
it does need to be satisfied once a region of allowed 
values for $k$ is known, which will be ascertained in 
the following. 

Concerning Eq.~(\ref{B5}), we are led to
\begin{widetext}
\begin{equation}
\frac{d R(2\de +1)(r_+ r_-)^{\de}}{\left(\frac{r_+ r_-}{R} + 
(1-k^2)R\right)^{\al}} \leq 
\frac{r_+^2 r_-^2(3\al + 1) + 2(1-k)r_+ r_-R^2(2\al (k-1) + 
2k -1)-(1-k)^3 R^4(k(\al+3) - \al + 3)}{\left[(1-k)^2 R^2 - 
r_+ r_-\right]\left[(k-1) R^2 (k(\al+3) - \al + 3)\right]},
\end{equation}
\end{widetext}
which does not contain any new information. 

On the other hand, 
when Eq.~(\ref{B6}) is simplified to
\begin{align}
\frac{d R(2\de +1)(r_+ r_-)^{\de}}{\left(\frac{r_+ r_-}{R} + 
(1-k^2)R\right)^{\al}} & \nonumber  \\
& \hspace{-15mm} \geq \frac{R^2(1-k^2)+(2 \al + 1)r_+ r_- 
- 2R^2 k^2 \al }{R^2(1-k^2)+r_+ r_- -2R^2k^2\al}\,,
\end{align}
and one notices that the numerator on the right side 
must be positive, another constraint on $k$ naturally appears, 
namely
\begin{equation}\label{k6s}
k \leq \sqrt{\frac{1}{2\al +1} + \frac{r_+ r_-}{R^2}}.
\end{equation}

Finally, the last condition (\ref{B7}) gives the inequality
\begin{equation}
r_+ r_-(\al+1) - R^2 \left[(\al+1)k^2 + \al -1\right] \geq 0
\end{equation}
which constricts the values of $k$ to be within the interval
\begin{equation}\label{k7s}
k \leq \sqrt{-\frac{\al -1}{\al +1} + \frac{r_+ r_-}{R^2}}.
\end{equation}

The definitive region of permitted values for $k$ is 
the intersection of the conditions (\ref{k1s}), (\ref{k2s}), 
(\ref{k6s}) and (\ref{k7s}). It is possible to show that 
such an intersection gives the range
\begin{equation}\label{intK}
\frac{\al}{\al+3} - \sqrt{\frac{9}{(\al+3)^2} + \frac{r_+ r_-}{R^2}} 
\leq k \leq \sqrt{-\frac{\al -1}{\al +1} + \frac{r_+ r_-}{R^2}}
\end{equation}
where $\al$ must be restricted to
\begin{equation}\label{intAl}
\al \geq \frac{1+\frac{r_+ r_-}{R^2}}{1-\frac{r_+ r_-}{R^2}}.
\end{equation}
Returning to Eq.~(\ref{B4s}), it is now possible to verify if 
the interval (\ref{intK}) satisfies said condition, 
which indeed it does.

\subsection{The black hole limit}
\label{bhl2}
 
If one takes the shell to its own gravitational radius, 
the chosen temperature equation of state (\ref{bF6})
is wiped out,
and a new 
equation of state 
sets in  to adapt to 
the quantum spacetime properties.
The new equation of state 
is then given by Eq.~(\ref{thawk})
and the black hole entropy 
(\ref{SS}) follows.

\section{Other equations of state}
\label{other}

Naturally, other equations of state can be sough.
We give four examples, one fixing  $b(r_+,r_-)$
and three others 
fixing $c(r_+,r_-)$.

If we fix 
the inverse temperature 
\begin{equation}
b(r_+,r_-) = \gamma\,
\frac{r_+^2}{r_+-r_-}\,,
\end{equation}
for some $\gamma$, as we did before, 
then generically, from Eq.~(\ref{bc5}), we find 
\begin{equation}
c(r_+,r_-) = \frac{a(r_+r_-)(r_+-r_-)+r_-}{r_+^2} \,,
\end{equation}
where $a(r_+r_-)$ is an arbitrary function
of integration
of the product $r_+r_-$
and presumably 
also depends on the intrinsic constants
of the matter that makes up the shell. 
Then, from Eq.~(\ref{dSQ}), the entropy is
\begin{equation}
S(r_+,r_-)=\frac{\gamma}{4G}\left(r_+^2 + 
\int_{0}^{r_{+}r_{-}}
\left(1-a(x)\right)\,dx\right)\,,
\end{equation}
where again we are assuming zero entropy 
when $r_+=0$.
In the example we gave previously we have 
put $a(r_+r_-)=1$, so that 
$c(r_+,r_-) = \frac{1}{r_+}$. This case 
$a(r_+r_-)=1$ gives precisely that the 
entropy of the shell 
is proportional to the area 
of its gravitational radius
and for $\gamma= \frac{4\pi}{\hbar}$
gives  that the 
entropy of the shell is equal to 
the corresponding black hole entropy
as we have discussed previously.
Of course, many 
other choices can be given
for $a(r_+r_-)$ and quite 
generally the entropy 
will be a function of $r_+$ 
and $r_-$, $S=S(r_+,r_-)$.

Inversely, instead of $b(r_+,r_-)$ one can give $c(r_+,r_-)$.
One equation for $c(r_+,r_-)$ could be 
\begin{equation}
c(r_+,r_-) =\frac{1}{r_{+}}\,,
\end{equation}
as for the black hole case. The 
integrability condition, Eq.~(\ref{bc5}),
for the temperature
then gives 
\begin{equation}
b(r_+,r_-)=\frac{h(r_{+})}{r_{+}-r_{-}}\,,
\label{neweos1}
\end{equation}
where $h(r_+)$ is a function that can be 
fixed in accord with the matter
properties of the shell.
Then, from Eq.~(\ref{dSQ}), the entropy is
\begin{equation}
S(r_+)=\frac{1}{2G}\int_{0}^{r_{+}}\,
\frac{h(x)}{x}\,dx\,,
\end{equation}
where it is implied that the function $h(x)$ vanishes at $x=0$ rapidly
enough so that the entropy goes to zero when $r_+=0$.
If we choose $h(r_+)=\frac{4\pi}{\hbar}
r_+^2$, then one recovers 
the black hole temperature and the black hole 
entropy for the shell.

Another equation of state one can choose
for $c(r_+,r_-)$ is
\begin{equation}
c(r_+,r_-) =\frac{1}{r_{-}}\,.
\end{equation}
The integrability condition, Eq.~(\ref{bc5}), similarly 
gives
\begin{equation}
b(r_+,r_-) =\frac{h(r_{-})}{r_{+}-r_{-}}\,.
\end{equation}
where $h(r_-)$ is a function that can be 
fixed in accord with the matter
properties of the shell.
In this case, from Eq.~(\ref{dSQ}),  the entropy 
of the shell depends on $r_{-}$ only,
and is given by
\begin{equation}
S(r_-)=\frac{1}{2G}\int_{0}^{r_{-}}\frac{h(x)}{x}\,dx\,,
\end{equation}
where we are assuming zero entropy 
when $r_-=0$.

Yet another example can be obtained if one puts
\begin{equation}
c(r_+,r_-) =c(r_{+}r_{-})\,,
\end{equation}
i.e., $c$ is a function of the product
$r_{+}r_{-}$
and may also depend on the
intrinsic constants of the matter that makes up the shell.
The integrability condition  
then gives
\begin{equation}
b=b_{0}\,,
\end{equation}
where $b_0$ is a constant, and so
in this case, the temperature measured at infinity 
does not depend on
$r_{+}$ or $r_{-}$.  The entropy is then
\begin{equation}
S(r_+,r_-)
=\frac{b_{0}}{2G}
\left(r_{+}+r_{-}-\int_{0}^{r_{+}r_{-}} c(x)\,dx\right)\,,
\end{equation}
where we are assuming zero entropy 
when $r_+=0$ and $r_-=0$.

One could study in detail these four cases
for the thermodynamics of a shell
performing 
in addition a stability analysis for each one. 
We refrain here to do so. Certainly other interesting
cases can be thought of.

\section{Conclusions}
\label{conc}

We have considered the thermodynamics of a self-gravitating
electrically charged thin shell thus generalizing previous works on
the 
thermodynamics of self-gravitating thin-shell systems.  Relatively to
the simplest shell where there are two independent thermodynamic state
variables, namely, the rest mass $M$ and the size $R$ of the shell, we
have now a new independent state variable in the thermodynamic system,
the electric charge $Q$, out of which, using the first law of
thermodynamics and the equations of state one can construct the
entropy of the shell $S(M,R,Q)$.  Due to the additional variable, the
charge $Q$, the calculations are somewhat more complex. Concomitantly,
the richness in physical results increases in the same proportion.

The equations of state one has to give are the 
pressure $p(M,R,Q)$, the 
temperature $T(M,R,Q)$,
and the electric potential $\Phi(M,R,Q)$.
The pressure can be obtained from dynamics alone, using the thin-shell
formalism and the junction conditions for a flat interior and a
Reissner-Nordstr\"om exterior.  The form of the temperature and of the
thermodynamic electric potential are obtained using the integrability
conditions that follow from the first law of thermodynamics.

The differential for the entropy in its final form shows 
remarkably
that the entropy must be a function of $r_+$ and $r_-$ alone, 
i.e., a function of the 
intrinsic properties of the shell spacetime.  
Thus, shells with the same $r_+$ and $r_-$ 
(i.e., the same ADM mass $m$ and charge $Q$) 
but different radii $R$, have the
same entropy. From the thermodynamics 
properties alone of the shell one cannot
distinguish a shell near its own gravitational 
radius from a shell far from it.
In a sense, the shell can mimic a black 
hole.

The differential for the entropy in its final form
gives that $T$ and $\Phi$
are related through an integrability condition.  One has 
then to
specify either $T$ or $\Phi$ and the form of the other 
function is somewhat constrained. 
We gave two example
cases and mentioned other possibilities. 

First, we gave the equations of state where the temperature has
the form of the Hawking temperature, apart from a constant factor, and
the electric potential has a simple precise form $Q/r_+$, and 
found the entropy.  When the factor is the Hawking factor it was shown
that the resulting entropy was equal to the Bekenstein-Hawking entropy
of a nonextremal charged black hole.  
The need to set the temperature of the shell 
equal to the Hawking temperature is 
justified when the shell is taken to its
own gravitational radius. 
At this radius the backreaction of the nearby quantum fields
diverges unless the shell has precisely the Hawking temperature.
Conversely, one
should note that if instead, the function for the
electric potential $Q/r_+$
was given, the integrability equation would then fix the function $T$
apart from an arbitrary function. 
A simple choice for this arbitrary function 
is the Hawking temperature.

Second, the other set of 
equations of state were given as 
a simple ansatz.
For the thermal equation of state, we set the temperature 
as proportional to some power in the ADM mass $m$, 
and the thermodynamic electric potential was set
to be a power in the electric charge and an 
inverse power in $m$.
This choice also allows one to find 
an expression
for the entropy of the shell and, furthermore,
allows for
an analytic stability analysis.
Indeed, despite the increase in complexity in
the thermodynamic stability analysis due to the existence of four new
stability equations, it was possible to obtain a unique range for
the redshift parameter
$k$, as well as the regions of allowed values for the parameters $\al$
and $\de$.

Many other interesting 
equations of state can be chosen and some of 
them were indeed given.  However at the
gravitational radius all turn into the Hawking
equation of state, i.e., the Hawking temperature.  
Since the area of the shell $A$
is equal to the gravitational radius area $A_+$, 
$A=A_+$, when the shell is at its own gravitational
radius, and $S=\frac{1}{4}\frac{A_+}{A_p}$
in this limit, 
we conclude that
the entropy of the shell is proportional to its
own area $A$. This indicates
that all its fundamental degrees of freedom
have been excited. 
Matter systems at their own gravitational radius are
called quasiblack holes and have thermodynamic 
properties similar to black
holes.

\begin{acknowledgments}
We thank FCT-Portugal for financial support through Project
No.~PEst-OE/FIS/UI0099/2013. GQ 
also ack\-now\-ledges the grant No.~SFRH/BD/92583/2013 from FCT.
OBZ has been partially supported by the Kazan Federal University 
through a state grant for scientific activities.

\end{acknowledgments}

\appendix \section{The dominant energy condition}
\label{apa}

With the expressions for the mass density and pressure, 
Eqs.~(\ref{Sig1})-(\ref{pQk1}), we
can consider some mechanical constraints which the shell should
naturally obey.  
One can impose that the shell satisfies
the weak energy condition.
It requires that $\sigma$ and $p$ be
positive, which is always verified.
One can also insist that the shell satisfies the 
dominant energy condition, i.e., 
\begin{equation}
p\leq \sigma\,. 
\end{equation}
It is then possible to show that the 
dominant energy
condition imposes the constraint $k \in [k_1,k_2]$, where
\begin{equation}
k_{1} = \frac{3}{5}\left(1 - \sqrt{1-
\frac{5}{9}\left(1-\frac{r_+ \, r_-}{R^2}\right)}\,\right)\,,
\end{equation}
and $k_{2} = \frac{3}{5}\left(1 + \sqrt{1-
\frac{5}{9}\left(1-\frac{r_+ \, r-}{R^2}\right)}\,\right)$.
Since $k_2>1$, and $k$ trivially obeys $k\leq1$,
we conclude that the dominant
energy condition restricts
the values of $k$ to obey 
\begin{equation}\label{d1}
k_1\leq k\,. 
\end{equation}
In the case where
there is no charge, i.e., $Q=0$ or $r_-=0$, 
one gets $k_1=1/5$, thus regaining the result
obtained in \cite{Mart}. When expressed in terms of the 
variables $R/m$ and $R/Q$, the relation (\ref{d1})
can be written as
\begin{equation}
\frac{R}{m}
\geq \frac{25}{6+10 \frac{G Q^2}{R^2} +3\sqrt{4+5\frac{G Q^2}{R^2}}}
\end{equation}
or in terms of $R/r_+$ and $r_-/R$,
\begin{equation}
\frac{R}{r_+} \geq
\frac{
12\sqrt{1- \frac{r_-}{R} - \left(\frac{r_-}{R}\right)^2 +
\left(\frac{r_-}{R}\right)^3}
+31 \frac{r_-}{R} -20 \left(\frac{r_-}{R}\right)^2 -12}
{24 \frac{r_-}{R} - 25
\left(\frac{r_-}{R}\right)^2}
\end{equation}
This is a mechanical constraint. 
A fundamental constraint, the no-trapped-surface 
condition for the shell, is $R\geq r_+$, as was given in 
Eq.~(\ref{notrapped}).

\section{Derivation of the equations of thermodynamic stability for a
system with three independent variables} 
\label{apb}

In this appendix we shall show the derivation of the equations of
thermodynamic stability for an electrically charged system,  
i.e., Eqs.~(\ref{B1})-(\ref{B7}).
Thus the approach used for two independent 
variables in \cite{callen} is extended here by us 
to three independent variables. We name 
these independent variables $M$, $A$, and $Q$.

We start by considering two identical subsystems, each with an entropy
$S = S(M,A,Q)$, where $M$ is the internal energy of the system
(equivalent to the rest mass), $A$ is its area and $Q$ its electric
charge. The usual state variables of a thermodynamic system are the
internal energy $U$, volume $V$ and the number of particles,
$N$, say. However, the system we wish
to study is an electrically 
charged thin shell, and thus it is natural to use the
variables $(M,A,Q)$. Thermodynamic stability is guaranteed if $dS = 0$
and $d^2 S < 0$ are both satisfied, or in other words, if the entropy
is an extremum and a maximum respectively.

Now suppose we keep $A$ and $Q$ constant and remove a positive amount
of internal energy $\Delta M$ from one subsystem to the other. The
total entropy of the two subsystems goes from the value $2 S(M,A,Q)$
to $S(M+\Delta M, A, Q) + S(M-\Delta M,A,Q)$. If the initial entropy
$S(M,A,Q)$ is a maximum, then the sum of initial entropies must be
greater or equal to the sum of final entropies, i.e.
\begin{equation}\label{Beq1}
S(M+\Delta M, A, Q) + S(M-\Delta M,A,Q) \leq 2 S(M,A,Q).
\end{equation}
Expanding $S(M+\Delta M, A, Q)$ and $S(M-\Delta M, A, Q)$ in a Taylor 
series to second order in $\Delta M$, we see that Eq.~(\ref{Beq1}) becomes
\begin{equation}\label{Beq2}
\left(\frac{\dd^2 S}{\dd M^2}\right)_{A,Q} \leq 0
\end{equation}
in the limit $\Delta M \to 0$. The same reasoning applies if we fix
$M$ and $Q$ instead and apply a positive change of area $\Delta A$, so
we must have
\begin{equation}\label{Beq3}
S(M, A+\Delta A, Q) + S(M,A-\Delta A,Q) \leq 2 S(M,A,Q).
\end{equation}
which in the limit $\Delta A \to 0$ gives
\begin{equation}\label{Beq4}
\left(\frac{\dd^2 S}{\dd A^2}\right)_{M,Q} \leq 0.
\end{equation} 
If we fix $M$ and $A$ and make a positive change $\Delta Q$ on the
charge, we have
\begin{equation}\label{Beq5}
S(M, A, Q+\Delta Q) + S(M,A,Q-\Delta Q) \leq 2 S(M,A,Q).
\end{equation}
and so it follows that
\begin{equation}\label{Beq6}
\left(\frac{\dd^2 S}{\dd Q^2}\right)_{M,A} \leq 0.
\end{equation} 
However, if we keep only one quantity fixed, like $Q$ for example, we
must also have a final sum of entropies smaller than the initial sum
if we apply a simultaneous change of area and internal energy rather
than separately, i.e.
\begin{align}\label{Beq7}
& S(M+\Delta M, A+\Delta A, Q)  \nonumber \\
& \hspace{4mm} + S(M-\Delta M,A-\Delta A,Q) \leq 2 S(M,A,Q).
\end{align}
This inequality is satisfied by Eq.~(\ref{Beq2}) and Eq.~(\ref{Beq4}),
but it also implies a new requirement. If we expand the left side in a
Taylor series to second order in $\Delta M$ and $\Delta A$, and use
the abbreviated notation $S_{ij} = \dd^2 S/\dd x_i \dd x_j$, we get
\begin{equation}
S_{MM} (\Delta M)^2 + 2 S_{MA} \Delta M \Delta A + 
S_{AA} (\Delta A)^2 \leq 0.
\label{smmsmasaa}
\end{equation}
Multiplying Eq.~(\ref{smmsmasaa}) by $S_{MM}$ and adding and
subtracting $S_{MA}^2 (\Delta A)^2$ to and from the left side, allows
the last inequality to be written in the form
\begin{equation}
(S_{MM} \Delta M + S_{MA} \Delta A)^2 + (S_{MM}S_{AA} -
S_{MA}^2)(\Delta A)^2 \geq 0.
\end{equation}
Since the first term on the left side is always greater than zero, we
see that it is sufficient to have
\begin{equation}\label{Beq8}
\left(\frac{\dd^2 S}{\dd M^2}\right) \left(\frac{\dd^2 S}{\dd
A^2}\right) - \left(\frac{\dd^2 S}{\dd M \dd A}\right)^2 \geq 0.
\end{equation}
This concludes the derivation of Eqs.~(\ref{B1}), (\ref{B2}) and (\ref{B4}). 

To derive 
the other stability equations, namely, 
Eqs.~(\ref{B3}), (\ref{B5}), (\ref{B6}), 
and (\ref{B7}), 
we note that
we can repeat the same calculations but now 
we fix $M$ and $A$ in
turns. It is now straightforward to see that, when fixing $M$, we must
have
\begin{equation}
S_{AA} (\Delta A)^2 + 2 S_{AQ} \Delta A \Delta Q + S_{QQ} (\Delta Q)^2 \leq 0,
\end{equation}
which is satisfied by
\begin{equation}\label{Beq9}
\left(\frac{\dd^2 S}{\dd A^2}\right) \left(\frac{\dd^2 S}{\dd
Q^2}\right) - \left(\frac{\dd^2 S}{\dd A \dd Q}\right)^2 \geq 0.
\end{equation}
Finally, by fixing $A$ we get the inequality
\begin{equation}
S_{MM} (\Delta M)^2 + 2 S_{MQ} \Delta M \Delta Q + S_{QQ} (\Delta Q)^2 \leq 0
\end{equation}
which implies the sufficient condition
\begin{equation}\label{Beq10}
\left(\frac{\dd^2 S}{\dd M^2}\right) \left(\frac{\dd^2 S}{\dd
Q^2}\right) - \left(\frac{\dd^2 S}{\dd M \dd Q}\right)^2 \geq 0.
\end{equation}
The last case left consists of doing a simultaneous change in all the
state variables of the system, i.e.,
\begin{align}\label{Beq11}
& S(M+\Delta M, A+\Delta A, Q+\Delta Q) \nonumber \\
& \hspace{4mm} + S(M-\Delta M,A-\Delta A,Q-\Delta Q) \leq 2 S(M,A,Q).
\end{align}
To investigate the sufficient differential condition that this
inequality implies, one must first expand $S(M+\Delta M, A+\Delta A,
Q+\Delta Q)$ and $S(M-\Delta M,A-\Delta A,Q-\Delta Q)$ in a Taylor
series to second order in $\Delta M$, $\Delta A$ and $\Delta Q$, which
can be shown to lead to
\begin{align}\label{Beq12}
& S_{MM} (\Delta M)^2 + S_{AA} (\Delta A)^2 + S_{QQ} (\Delta Q)^2
\nonumber \\ & + 2 S_{MA} \Delta M \Delta A + 2 S_{MQ} \Delta M \Delta
Q + 2 S_{QA} \Delta A \Delta Q \leq 0.
\end{align}
Multiplying the above relation by $S_{MM}$, noting that
\begin{align}
& (S_{MM} \Delta M + S_{MA} \Delta A + S_{MQ} \Delta Q)^2 = \nonumber
\\ & S_{MM}^2 (\Delta M)^2 + S_{MA}^2 (\Delta A)^2 + S_{MQ}^2 (\Delta
Q)^2 + \nonumber \\
& + 2 S_{MM} S_{MA} \Delta M \Delta A + 2 S_{MM} S_{MQ} \Delta M
\Delta Q + \nonumber \\
& + 
2 S_{MA} S_{MQ} \Delta A \Delta Q \,,
\end{align}
and inserting this into Eq.~(\ref{Beq12}), gives
\begin{align}
& (S_{MM} \Delta M + S_{MA} \Delta A + S_{MQ} \Delta Q)^2 \nonumber \\
& \hspace{1mm} + (S_{MM}S_{AA} - S_{MA}^2)(\Delta A)^2 + 
\nonumber \\
& \hspace{1mm} +
(S_{MM}S_{QQ}
- S_{MQ}^2)(\Delta Q)^2 +\nonumber \\
& + 2 (S_{MM}S_{QA}-S_{MA}S_{MQ}) \Delta A \Delta Q \geq 0.
\end{align}
Recalling Eq.~(\ref{Beq8}) and Eq.~(\ref{Beq10}), and noting that the
first term in the above inequality is always positive, we conclude
that the condition
\begin{equation}
\left(\frac{\dd^2 S}{\dd M^2}\right) \left(\frac{\dd^2 S}{\dd Q \dd
A}\right) - \left(\frac{\dd^2 S}{\dd M \dd A}\right) \left(\frac{\dd^2
S}{\dd M \dd Q}\right) \geq 0
\end{equation}
is sufficient to satisfy Eq.~(\ref{Beq11}). This concludes the
derivation of Eqs.~(\ref{B3}), (\ref{B5}), (\ref{B6}) and (\ref{B7}).
Thus all stability equations, Eqs.~(\ref{B1})-(\ref{B7}), have been
derived.

\end{document}